\documentclass[11pt,twoside,center]{article}
\usepackage{graphicx}

\usepackage{amsmath}
\usepackage{amsthm}
\usepackage{amsfonts}
\usepackage{amssymb}
\usepackage[english]{babel}

\newcommand{\be}{\begin{equation}}
\newcommand{\ee}{\end{equation}}
\newcommand{\bea}{\begin{eqnarray}}
\newcommand{\eea}{\end{eqnarray}}
\newcommand{\di}{\displaystyle}

\newtheorem{definition}{Definition}[section]

\title{Phase Diagram for Roegenian Economics}

\author{C. Udriste, M. Ferrara, I. Tevy, \\D. Zugravescu, F. Munteanu\footnote{
 Prof. Emeritus Dr. Constantin Udriste, Prof. Dr. Ionel Tevy, University  Politehnica of Bucharest, Faculty of Applied Sciences, Department of Mathematics-Informatics, Splaiul Independentei 313, Bucharest, 060042, Romania; {\tt udriste@mathem.pub.ro , \tt vascatevy@yahoo.com}
\newline Prof. Dr. Massimiliano Ferrara, Di.Gi.ES, University Mediterranea of Reggio Calabria,
Decisions LAB, Cittadella universitaria, seconda Torre, Loc. Feo di Vito, 89125 Reggio Calabria, Italy;
 {\tt massimiliano.ferrara@unirc.it}
\newline Acad. Dr. Dorel Zugravescu, Associate Prof. Dr. Florin Munteanu, Institute of Geodynamics "Sabba S. Stefanescu", Dr. Gerota 19-21,  Bucharest, 020032, Romania; {\tt dorezugr@geodin.ro , \tt florin@geodin.ro} }}
\date{Centenary of Romanian Great Union 1818-2018}
\begin{document}

\maketitle

\pagestyle{myheadings}
\markboth{Udriste, Ferrara, Tevy, Zugravescu, Munteanu}{Phase Diagram for Roegenian Economics}

\bigskip

\begin{abstract}
We recall the similarities between the concepts and techniques of Thermodynamics and Roegenian Economics.
The Phase Diagram for a Roegenian economic system highlights a triple point
and a critical point, with related explanations.
These ideas can be used to improve our knowledge and understanding of the nature of
development and evolution of Roegenian economic systems.
\end{abstract}

{\bf AMS Mathematical Classification}: 80A99, 91B54, 91B74

{\bf J.E.L. Classification: B41}

{\bf Key words}: Thermodynamic-economic dictionary, Roegenian economics,
economic triple point, economic critical point.

\section{Thermodynamic-Economic Dictionary}

Thermodynamics is important as a model of phenomenological theory, which describes and
unifies certain properties of different types of physical systems. There are many
systems in biology, economics and computer science, for which an organization
similar and unitary-phenomenological would be desirable. Our purpose is to present
certain features of the economy that are inspired by thermodynamics and vice versa. In this context, we offer a Dictionary that reflects the Thermodynamics-Economy isomorphism. The formal analytical-mathematical analogy between economics and thermodynamics is now well-known or at least accepted by economists and physicists as well (see also Econophysics \cite{[1]} which is a discipline in this sense).

Starting from these observations, the works of Udriste et al. \cite{[11]}-\cite{[19]} build an isomorphism between thermodynamics and the economy, admitting that fundamental laws are also in correspondence through our identification. Therefore, each thermodynamic system is naturally equivalent to an economic system, and thermodynamic laws have correspondence in the economy.

In the following, we reproduce the correspondence between the characteristic state variables
and the laws of thermodynamics with the macro-economics as described in
Udriste et al. \cite{[11]}-\cite{[19]} based on the theory on which the
Roegenian economy is founded in 1971 \cite{[3]}. We do not know if
Georgescu-Roegen would have judged this, but that's why we are judging him instead.

{\it The Gibbs-Pfaff fundamental equation in thermodynamics} $ dU-TdS + PdV + \sum_k \mu_k dN_k = 0 $ is changed to
{\ Gibbs-Pfaff fundamental equation of economy} $ dG-IdE + PdQ + \sum_k \nu_k d {\cal {N}} _ k = 0 $. These equations are combinations of the first law and the second law (in thermodynamics and economy respectively). The third law of thermodynamics $ \di \lim_{T \to 0} S = 0 $ suggests the third law of economy $ \di \lim_ {I \to 0} E = 0 $ "if the internal political stability $ I $ tends to $ 0 $, the system is blocked, meaning entropy becomes $ E = 0 $, equivalent to maintaining the functionality of the economic system must cause disruption").

Process variables $ W = $ {\it mechanical works} and $ Q $ = {\it heat} are introduced into elementary mechanical thermodynamics by $ dW = PdV $ (the first law) and by elementary heat, respectively, $ dQ = TdS $, for reversible processes,
or $ dQ <TdS $, for irreversible processes ({\it second law}).
Their correspondence in the economy, $$ W = \hbox {\it wealth of the system},\,\,q= \hbox{\it production of goods}$$
are defined by $ dW = Pdq $ ({\it elementary wealth in the economy}) and $ dq = IdE $ or $ dq <IdE $ ({\it the second law or the elementary production of commodities}). A commodity is an economic good,
a product of human labor, with a utility in the sense of life,
for sale-purchase on the market in the economy.

Sometimes a thermodynamic system is found in an
{\it external electromagnetic field} $ (\vec {E}, \vec {H}) $. {\it The external electric field} $ \vec E $ determines
the {\it polarization} $ \vec P $ and {\it the external magnetic field} $ \vec H $ determine {\it magnetizing} $ \vec M $. Together they give the total elementary mechanical work  $ dW = PdV + \vec E d \vec P + \vec H d \vec M $. Naturally, an economic system is found in an {\it external econo-electromagnetic field} $(\vec {e}, \vec {h})$. The {\it external investment (econo-electric) field} $\vec e$ determines
{\it initial growth condition field (econo-polarization field)} $ \vec p $
and the {\it external growth field (econo-magnetic field)} $ \vec h $
cause {\it growth (econo-magnetization)} $ \vec m $. All these fields produce the elementary mechanical work $dW=PdQ+\vec e
d\vec e +\vec h d\vec m$. The economic fields introduced here are imposed on the one hand by the type of economic system and on the other hand by the policy makers (government, public companies, private firms, etc.).

\vspace{1cm}
{\hspace{-1cm}
\begin{tabular}{lll}
\vspace{0.3cm}
\hspace{0.5cm}THERMODYNAMICS&\hfill & \hspace{0.7cm}ECONOMICS \\
U=\hbox{internal energy} & \hfill\ldots &G=\hbox{growth potential}\\
T=\hbox{temperature}&\hfill\ldots& I=\hbox{internal politics stability}\\
S=\hbox{entropy}&\hfill \ldots& E=\hbox{entropy}\\
P=\hbox{pressure}& \hfill \ldots& P=\hbox{price level (inflation)}\\
V=\hbox{volume}&\hfill \ldots& Q=\hbox{volume, structure, quality}\\
M=\hbox{total energy (mass)}&\hfill \ldots& Y=\hbox{national income (income)}\\
Q=\hbox{electric charge}& \hfill \ldots& $\mathcal{I}$=\hbox{total investment}\\
J= \hbox{angular momentum}&\hfill \ldots& J=\hbox{economic angular momentum}\\
\hspace{1cm}\hbox{(spin)}&\hfill &\hspace{1cm}\hbox{(economic spin)}\\
M=M(S,Q,J)& \hfill \ldots& Y=Y(E,{$\mathcal{I}$},J)\\
$\Omega = \frac{\partial M}{\partial J}$= \hbox{angular speed}&\hfill \ldots& $\frac{\partial Y}{\partial J}$=\hbox{marginal inclination to rotate}\\
$\Phi = \frac{\partial M}{\partial Q}$=\hbox{electric potential}&\hfill \ldots& $\frac{\partial Y}{\partial {\cal{I}}}$=\hbox{marginal inclination to investment}\\
$T_H= \frac{\partial M}{\partial S}$=\hbox{Hawking temperature}&\hfill \ldots& $\frac{\partial Y}{\partial E}$=\hbox{marginal inclination to entropy}\\
$G$ = Newton constant &\hfill \ldots& $\mathcal{G}$= universal economic constant\\
$c$ = light velocity &\hfill \ldots& $c$= maximum universal exchange speed \\
$\hbar$= normalized Planck constant &\hfill \ldots& $\hbar$ = normalized economic quantum \\
\end{tabular}}
\vspace{1cm}

The long term association between Economy and Thermodynamics can be strengthened
with new tools based on the previous dictionary. Of course, this new idea of
the thermodynamically-economical dictionary produces concepts different from those in econophysics.
Econophysics \cite{[1]} seems to build similar economic notions to physics, as if those in the economy
were not enough.

Our thermodynamic-economic dictionary allows the transfer of
information from one discipline to another,
keeping the background of each discipline, that we think that was suggested by Georgescu-Roegen \cite{[3]}.
Considering the far-fetched and mechanical classical liberal economy, Georgescu-Roegen pointed out
the contradiction between the second principle of thermodynamics and the law of entropy - that is,
between the unavoidable degradation of natural resources used by mankind as a result of their
use and unlimited material growth. He showed the adherence to an economic decline to
take into account the physical law of entropy.

A complementary point of view can be found in \cite{[7]}, \cite{[9]}. The papers \cite{[4]},
\cite{[41]}, \cite{[8]} completes our thoughts.

\begin{definition}
{\it An economy structured similar to thermodynamics is called Roegenian economy}.
\end{definition}

In the economic field, the means of thermodynamics are preferable to mechanical ones.
The macroscopic thermodynamic quantities are not true point-wise mathematical functions.
They are obtained through statistical mediation. Also in the economy it is
impossible point-wise prediction.

\section{Triple Point in Economic Phase Diagram}

An economic system is a system of production, resource allocation and distribution
of goods and services within a society or a given geographic area. It includes the
combination of the various institutions, agencies, entities, decision-making processes
and patterns of consumption that comprise the economic structure of a given community.
As such, an economic system is a type of social system. The mode of production is a related concept.
All economic systems have three basic questions to ask: what to produce, how to produce and in
what quantities and who receives the output of production?

\begin{figure}
  \centering
  \includegraphics[width=9cm]{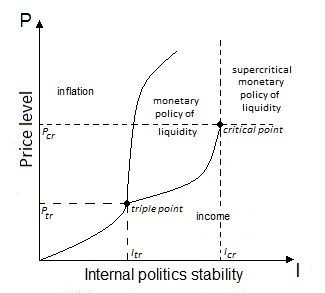}\\
  \caption{Triple point, Critical point}\label{Fig.1}
\end{figure}

The Figure $1$ gives an example of a phase diagram that focuses the effect
of the internal politics stability and the level prices in an economic system,
inspired by thermodynamics \cite{[JV]}, via the dictionary in Section $1$, completed in this Section.
According this dictionary, the "pressure" is replaced by "price level" and "temperature"
is replaced by "internal politics stability". Consequently,
each point in the diagram represents a possible combination between the internal politics stability
and the level of prices for the studied economic phase.
The $2D$ chart contains three regions that represent the correlation between
different states of the economy: (1) inflation,  (2) monetary policy as liquidity, (3) income.
More precisely, we accept the following association chart

\vspace{1cm}
\begin{centerline}
{\hspace{-1cm}
\begin{tabular}{lll}
\vspace{0.3cm}
\hspace{0.5cm}SUBSTANCE\hspace{6cm}&\hfill\ldots\ldots & \hspace{-0.2cm}ECONOMIC SYSTEM \\
T=\hbox{temperature}\hspace{6cm}&\hfill\ldots\ldots& \hspace{-0.2cm} I=\hbox{internal politics stability}\\
P=\hbox{pressure}\hspace{6cm}& \hfill \ldots\ldots&\hspace{-0.2cm} P=\hbox{price level}\\
\hbox{solid}\hspace{6cm}&\hfill \ldots\ldots& \hspace{-0.1cm}\hbox{inflation}\\
\hbox{fluid}\hspace{6cm}& \hfill \ldots\ldots& \hspace{-0.1cm}\hbox{monetary policy as liquidity}\\
\hbox{gas}\hspace{6cm}&\hfill \ldots\ldots& \hspace{-0.1cm}\hbox{income}\\
\end{tabular}}
\end{centerline}
\vspace{1cm}

Phase diagram (Figure 1) is a graphical representation of the economic states of a Roegenian economical system
under different conditions of internal politics stability $I$ and price level $P$. Our
phase diagram has internal politics stability $I$ on the $x$-axis and price level $P$ on the $y$-axis.
As we cross the curves on the phase diagram, a phase change occurs. In addition,
two states of the economic system coexist in equilibrium on some significant curves.

{\bf The "apparently strange"} association between "substance and economic system"
is due to the signification of the horizontal axis (internal politics stability).

 We recall some important phases into a Roegenian economical system.

{\bf Inflation}: In economics, inflation is a sustained increase in the price level of goods
and services in an economy over a period of time. When the price level rises, each unit of
currency buys fewer goods and services; consequently, inflation reflects a reduction in
the purchasing power per unit of money – a loss of real value in the medium of exchange
and unit of account within the economy. A chief measure of price inflation is the
inflation rate.

{\bf Monetary policy as liquidity or consumption}: Monetary policy is the process by which the monetary authority of a country, typically the central bank or currency board, controls either the cost of very short-term borrowing or the monetary base, often targeting an inflation rate or interest rate to ensure price stability and general trust in the currency.

Consumption is the using of goods and services in an economy, or the amount of goods and services used.
Liquidity means how quickly you can get your hands on your cash.
In simpler terms, liquidity is to get your money whenever you need it.

{\bf Income}: In the field of public economics, the term income may refer to the accumulation of
both monetary and non-monetary consumption ability, with the former (monetary) being used as a proxy for total income.
Gross income can be defined as sum of all revenue. Net income = Revenue - Expenses.

In the following we assume that an economic system is found in one of the three previous states
(inflation, monetary policy as liquidity, income).

The stability in the domestic political and government actions are important contributions to
attract the foreign investors to invest in the country.  The foreign
investment can make economic growth and developing host countries.

The internal politics un-stability and the high level of prices favor training
inflation. Stability of increased domestic policy and low price levels generate income.
Monetary policy as liquidity lies between these extremes.

\begin{definition}
{\it The common point of the three curves, "inflation - monetary policy of liquidity" equilibrium (recovery-recession curve),
"monetary policy of liquidity - income" equilibrium (increasing-decreasing curve),
"inflation - income" equilibrium (economic boom-crisis curve), is called triple point}.
\end{definition}

Where all three curves meet, we must have a unique combination of
internal politics stability and price level where all three phases
are in equilibrium together.

Generally, a phase diagram in physical chemistry, engineering, mineralogy, materials,
and economics science is a type of 2D chart used to show that three states $(c_1, c_2, c_3)$
coexist and they are at equilibrium depending on two primary indicators $(p_1,p_2)$ on coordinate axes.

{\bf Related Questions} Which are the implications of the triple point on real economic system issues?
Do any economic system posses a triple point?

\section{Critical Point in Economic Phase Diagram}

Again, we relate to the similarity between a thermodynamic system  \cite{[JV]} and an economic system,
obtaining the 2D economic phase diagram in Figure 1.

Phase diagram illustrate the variations between the states of an economic system,
"inflation, monetary policy as liquidity, and income",
as they relate to internal politics stability and price level. The previous phase diagram highlights two important points:
{\bf triple point} - the point on a phase diagram at which the three states of an economic system, "inflation
monetary policy as liquidity, income", coexist;
{\bf critical point} - the point on a phase diagram at which the increasing-decreasing trend curve ends.
\begin{definition}
{\it The point $(I_c,P_c)$ from which, the conditions $P > P_c$ (high level prices), and
$I>I_c$ (high internal politics stability) shows that
price levels cancel economic growth and conversely, is called critical point}.
\end{definition}

Phase diagram plot price level (typically in reference currency) versus
internal politics stability (typically in percents - see, for instance, Moody's rating).
The labels on the graph represent the stable states of a system in equilibrium.
The curves represent the combinations of price level and internal politic stability
at which two phases can exist in equilibrium. In other words, these curves define
phase change points. The curve between the inflation and income phases,
represents economic boom (inflation to income) and crisis (income to inflation).
The curve which joins the triple point and the critical point
divides the income and monetary policy of liquidity phases and represents
economic increasing (monetary policy of liquidity to income), or economic
decreasing (income to monetary policy of liquidity).
The curve between the inflation and the monetary policy of liquidity phases,
represents recovery-recession curve (inflation to monetary policy of liquidity -
monetary policy of liquidity to inflation). There are also
two important points on the diagram, the triple point and the critical point.
The {\bf triple point} represents the combination of price level and internal politics stability
that facilitates all phases of an economic system at equilibrium. The {\bf critical point} is
the ending point of the "monetary policy of liquidity-income" phase curve and relates to the
critical price level, the price level above, which a supercritical monetary policy of liquidity forms.

The critical point $(I_{cr},P_{cr})$ is a projection of degenerate critical point
$(I_{cr}, P_{cr}, Q_{cr})$ of a constant level $I=const$ curve in the Van der Waals surface.

The triple point and the critical point reflect important changes in interior of economic $I-P$ diagram.
In Roegenian economics, the triple point of an economic system is a pair (internal politics stability,
price level) at which the three phases (inflation, monetary policy of liquidity, and income)
of that economic system coexist in economic equilibrium. Also, a critical point (or critical state)
is the end point of a phase equilibrium curve in a Roegenian economic system.

\section{Conclusion}

From the origins of the nineteenth century, the laws of thermodynamics
proved to be robust. We investigate deep connections
between thermodynamics and economic theory, connections that allow us
to emphasize that Georgescu-Roegen's prediction is full of content \cite{[DI]}.
These links will lead to common explanations for the phenomena of thermodynamics and economics,
sometimes different from those in econophysics.

Roegenian economics (also called bioeconomics by Georgescu-Roegen) is both a
transdisciplinary and an interdisciplinary field of academic research addressing
the interdependence and coevolution of human economies and natural ecosystems,
both intertemporally and spatially.

Moving about the phase diagram $I-P$ reveals information about the phases of
an economic system. This interpretation holds in economics
and comes from the thermodynamics - specific diagram. It was wonderful in finding
this parallelism between fields of Science. An unicum by which reveals a
file rouge unifying different kind of human interactions with the nature.
This is the great intuition of Georgescu-Roegen and for us was the
premium moves to develop this theory starting from this scientific platform.

This paper aims to show how the common notions of thermodynamics and Roegenian economics work.
Obviously, the similarities can not be total, but we propose a thermodynamic model
of the economy, the Roegenian economy. In future works we shall give examples of
economic systems characterized by the pairs (critical internal politics stability, critical price levels).

High - "internal politics stability" phase diagrams including the critical point of
income - "monetary policy as liquidity or consumption" phase transition as well as
price-level $(P,Q,I)$, production-level $(q,Q,I)$ and entropic $(E,Q,I)$ equations of states
are very important in many aspects of theoretical economics.

\end{document}